\documentclass[CRMECA,Unicode,manuscript]{cedram}
\newcommand{\bfu}{\mbox{\boldmath $u$}}

\newcommand{\bfx}{\mbox{\boldmath $x$}}
\newcommand{\bfy}{\mbox{\boldmath $y$}}
\newcommand{\bfz}{\mbox{\boldmath $z$}}
\newcommand{\eps}{{\epsilon}}

\newcommand{\lam}{{\lambda}}
\newcommand{\sig}{{\sigma}}

\newcommand{\Gam}{{\Gamma}}

\newcommand{\Ome}{{\Omega}}
\newcommand{\bfeps}{\mbox{\boldmath $\epsilon$}}

\newcommand{\bfsig}{\mbox{\boldmath $\sigma$}}

\graphicspath{{./fig/lip_figures_0d/},{./fig/}}

\newcommand{\dist}{\mathrm{dist}}

\newcommand{\rmd}{\ensuremath{\mathrm{d}}}
\newcommand{\dint}{\mbox{\, \rmd}}

\newcommand{\Yc}{Y_\mathrm{c}}
\newcommand{\lc}{l_\mathrm{c}}
\newcommand{\Gc}{G_\mathrm{c}}

\newcommand{\dotalp}{\dot{\alpha}}
\newcommand{\texton}{\text{\ on\ }}

\newcommand{\textif}{\text{\ if\ }}
\newcommand{\textor}{\text{\ or\ }}

\newcommand{\Fe}{F_\mathrm{e}}

\newcommand{\Fp}{F_\mathrm{p}}
\newcommand{\Fsp}{F_\mathrm{sp}}
\newcommand{\Fse}{F_\mathrm{se}}
\newcommand{\Fsep}{F_\mathrm{sep}}
\newcommand{\fe}{f_\mathrm{e}}

\newcommand{\fp}{f_\mathrm{p}}
\newcommand{\fsp}{f_\mathrm{sp}}
\newcommand{\fse}{f_\mathrm{se}}
\newcommand{\fsep}{f_\mathrm{sep}}
\newcommand{\fx}{f_\mathrm{x}}

\newcommand{\abs}[1]{\lvert #1 \rvert}

\newcommand{\lip}{\mathrm{lip}}

\newcommand{\dt}{\Delta t}
\newcommand{\od}{\overline{d}}

\newcommand{\ep}{\epsilon_p}
\newcommand{\sigy}{\sigma_\mathrm{y}}
\newcommand{\epsy}{\epsilon_\mathrm{y}}

\newcommand{\dn}{d_n}
\newcommand{\Dn}{D_n}

\newcommand{\piu}{\pi^\mathrm{u}}
\newcommand{\pil}{\pi^\mathrm{l}}

\newcommand{\ud}{u_{\mathrm{d}}}
\newcommand{\sigc}{\sig_{\mathrm{c}}}
\newcommand{\epsc}{\epsilon_{\mathrm{c}}}
\newcommand{\epsp}{\epsilon_{\mathrm{p}}}
\newcommand{\epspn}{\epsilon_{\mathrm{p}n}}
\newcommand{\epspi}{\epsilon_{\mathrm{p}i}}
\newcommand{\Lip}{\mathrm{L}}

\title{Lipschitz regularization  for softening material models: the Lip-field approach}
\alttitle{Régularisation de Lipschitz pour les modèles de matériaux adoucissants : l'approche Lip-field}

\author{\firstname{Nicolas} \lastname{Mo\"{e}s} \IsCorresp}
\address{Ecole Centrale de Nantes, GeM Institute, UMR CNRS 6183, 1 rue de la Noë, 44321 Nantes, France}
\address{Institut Universitaire de France (IUF)}
\email[N. Mo\"{e}s]{nicolas.moes@ec-nantes.fr}
\author{\firstname{Nicolas}  \lastname{Chevaugeon}}
\addressSameAs{1}{Ecole Centrale de Nantes, GeM Institute, UMR CNRS 6183, 1 rue de la Noë, 44321 Nantes, France}
\email[N. Chevaugeon]{nicolas.chevaugeon@ec-nantes.fr}

\begin{abstract}
	Softening material models are known to trigger spurious localizations.
	This may be shown theoretically by the existence of solutions 
	with zero dissipation when localization occurs and numerically with spurious mesh dependency and localization in a single layer of elements.
	We introduce in this paper a new way to avoid spurious localization.  
	The idea is to enforce  a Lipschitz  regularity on the  internal variables responsible for the material softening. The regularity constraint introduces the needed length scale  in the material formulation. 
	Moreover, we prove bounds on the domain affected by this constraint. A first one-dimensional finite element implementation is proposed for softening elasticity and softening plasticity. 
\end{abstract}

\begin{altabstract}
	Les modèles de matériaux adoucissants sont connus pour déclencher des localisations parasites.
	Cela peut être démontré théoriquement par l'existence de solutions avec une dissipation nulle lors de la localisation et numériquement avec une dépendance de maillage et une localisation  dans une seule couche d'éléments.
	Nous introduisons dans cet article une nouvelle façon d'éviter les localisations parasites.
	L'idée est d'imposer une régularité de Lipschitz sur les variables internes responsables de l'adoucissement. La contrainte de régularité introduit l'échelle de longueur nécessaire dans la formulation du matériau.
	De plus, nous prouvons des bornes sur le domaine affecté par cette contrainte. Une première mise en œuvre par éléments finis unidimensionnels est proposée pour l'adoucissement élastique ou plastique. 
\end{altabstract}


\altkeywords{adoucissement, localisation, 
	endommagement, plasticité, Lipschitz, Lip-field}

\begin{document}
	
\maketitle

\section{Introduction}
%\cite{Crandall2001}, \cite{Crandall83},
%\cite{Mumford1989}
%\cite{Ambrosio90} \cite{Ambrosio97} \cite{AmbrosioT90}
%\cite{Buliga1998a} \cite{Buliga2009}
%\cite{Mielke2005}
With softening, the stress that a material can sustain is diminishing 
as the strain increases. This phenomenon exists both for elasticity
and plasticity. For elasticity, the stiffness is decreasing as the strain 
increases whereas for plasticity, the yield stress is diminishing as the plastic
strain accumulates.
Dealing with these types of models in finite element analysis is a challenge.
From the mathematical point of view, these models loose the nice convex 
properties of classical elastic models  or hardening plasticity models. 
%Existence and uniqueness of solutions is no longer ensured.
Limit points may exist: the solution does not exist  
beyond some loads. Bifurcation points are also possible: 
at some stage in time, several solutions start to exist (stable or not).
Among the many solutions that may exist, some are called spurious localizations. In the one-dimensional setting, these localizations are characterized by a softening occuring in a single point.
For higher dimensions, the localization occurs on zero-measure domains.
As a consequence, the dissipation involved in these zones is zero.
In other words, the external energy or load needed to diminish the bearing capacity of the structure is highly under-estimated.
The mathematical difficulties of softening material models have a direct 
impact on their numerical treatment. Spurious mesh dependency is 
observed with finite elements: the mesh orientation  has a strong impact on the results and, as the mesh is refined, only a single  element or a layer of elements, depending on the problem dimension, is affected by localization.

This paper introduces a new way to 
eliminate these spurious localizations from the  model.  
It does not mean that the solution is now unique but, at least, these unwanted solutions are removed.
The design of approaches  to remove spurious localizations has been going on for 
about fourty years now.

Regarding quasi-static analysis of time-independent models, 
several remedies have been studied in the literature.
They all share in common the fact that
a regularizing length scale is injected in the model. 
For the so-called non-local integral damage model, the damage evolution at a given point is governed by a driving force which  is the average of the local driving force over some distance around that point \cite{Bazant84,Pijau87,Bazant2002,Lorentz2003a}. 
\textcolor{black}{Refinements of the non-local integral model with an evolving internal length may be found in \cite{Giry2011,Rastiello2018}}.

In higher order, kinematically based, gradient models, the length scale is introduced through 
the inclusion of higher order deformation gradient in the energy expression \cite{Aifantis84,Aifantis86,Schreyer86} or through additional rotational degrees of freedom \cite{Muhlhaus1987}.
For higher order, damage based, gradient models, the energy depends on 
the gradient of the damage  thus involving again a length scale
\cite{Fremond1996,Pijaudier-Cabot1996,Peerlings2001a,Nguyen2005b}.

\textcolor{black}{Regarding the energy minimizing approaches, they mainly  stem from the seminal paper by  Mumford and Shah \cite{Mumford1989}. The Italian school of calculus of variations has given most of the mathematical background for models of brittle fracture based on the Mumford-Shah function \cite{Ambrosio90,Ambrosio97}  and the variational approximations \cite{AmbrosioT1990}. A very early paper which uses minimizing movements \cite{Giorgi93} and variational approximations is \cite{Buliga1998a}  and in \cite{Buliga2009}, the Ambrosio Tortorelli approximation is used for a damage energy minimizing model, which turned out to be related, mathematically, with viscosity-based approaches to introduce a length scale into damage. The Ambrosio Tortorelli approximation  was also used by Bourdin et al \cite{Bourdin2000a} to implement the revisit of brittle fracture introduced in  \cite{Francfort1998}, leading to the so-called variational approach to fracture \cite{Mielke2005,Bourdin2008}.}

At about the same time, the phase-field approach was
emanating from the physics community \cite{Karma2001,Hakim2009} and  then 
developped for mechanics applications
\cite{Amor2009,Miehe2010,Miehe2010b,Kuhn2010,Ambati2015a}.

Yet another way to introduce a length scale is the Thick Level Set approach to fracture. The approach was introduced for brittle damage in \cite{Moes2011,Stolz2012}.
The damage evolution is tied to a distance field. 
The damage front on which the damage starts  is the level set zero and it grows to a value of 1
(fully damage state) at some distance $\lc$ in the wake of the front.
That amounts to imposing the norm of the damage gradient on zones with strictly positive damage (equality constraint). 
The possibility of an inequality constraint on the damage gradient was also considered in  \cite{Moes2014,Stershic2017} allowing to combine diffuse and localizing damage fields. 
The advantage of this inequality was further stressed in \cite{Fremond2017}, 
where it was shown that the inequality constraint is convex on the contrary to the equality
constraint. This paper also demonstrates that a level set field is not mandatory:
a variational approach with Lagrange multipliers enforcing the inequality constraint
may be used. Following the ideas of \cite{Fremond2017},  an implementation  is  provided  in  \cite{Nunziante2019}.

The Lipschitz regularization introduced in this paper enforces a 
regularity on the damage field. 
The obtained field is called a Lip-field.
The Lipschitz regularity does not require  the existence of the damage gradient. Yet, the Lip-field is  differentiable almost everywhere.
To present the Lip-field concept, the appealing framework 
of  incremental energetic variational  potential is considered.
The interest of this framework  was first demonstrated for visco-plasticity \cite{Mialon1986,Ortiz1999} and was later reused as a building block
for the variational approach to fracture \cite{Bourdin2008}.
When softening may occur, 
the incremental potential is not convex  
but only separately convex with respect to the
displacement (and non-softening internal variable) on one-side and
softening variable on the other side.
It is thus natural to proceed with an alternate minimization 
of successive convex problems \cite{Boyd2004}. The Lip-field convex constraint is added as an 
extra  constraint 
in the alternate minimization. 
We demonstrate in this paper upper and lower 
bounds for the minimization over the damage field. 
These bounds reduce drastically the zone over which the Lipschitz  condition needs to be activated.
The proof holds in any spatial dimension.

A one-dimensional finite element implementation is provided for the 
Lip-field approach.
The potential needs to be minimized for the nodal values of the displacement and for the  internal variables located at the integration point
in each finite element,  including the damage state variable.
In other words, with the Lip-field approach, the damage variable may be kept at the integration points with the other internal variables and does not
need to be stored at the nodes, thus  following 
the common practice for nonlinear finite element analysis.

Finally, note that there is also a current interest in Lipschitz regularization to improve the robustness to adversarial perturbations
for learning framework \cite{krishnan2020}.

The paper is organized as follows. 
The next section describes the classical mechanical formulation 
for non-softening material models. In Section \ref{sec:dam}, a softening variable is 
introduced and the Lipschitz constraint is imposed. Elastic and plastic softening models 
are presented in Section \ref{sec:mod} for the one-dimensional setting. 
Finite element analysis are carried out in Section \ref{sec:fem}. Discussion and 
future works are provided in Section \ref{sec:conc}.

%\section{Material model with Internal variables} \label{sec:class}
\section{Generalized standard materials} \label{sec:class}
We consider the deformation of a body initially 
occupying 
a domain $\Ome$ through a displacement field $\bfu$.
For simplicity, we assume small, quasi-static deformations. The Cauchy stress is denoted  $\bfsig$
and the strain  $\bfeps$
\begin{equation}\label{key}
	\bfeps(\bfu) = \frac{1}{2} (\nabla \bfu + (\nabla \bfu)^\mathrm{T}) 
\end{equation}
where $\nabla$ indicates the gradient operator.
Regarding the boundary conditions, 
the displacement is controlled on a part of the boundary denoted 
$\Gam_u$ assumed fixed in time. On the rest of boundary,  
zero traction force are assumed (again for simplicity).
To be kinematically admissible at some instant $t$, the displacement 
field must belong to $U(t)$:
\begin{equation}\label{eq:kin}
	U(t) = \{ \bfu \in H^1(\Ome): \bfu = \bfu_d(t) \texton \Gam_u \}
\end{equation}
%The assumptions above are not essential for the development 
%of the ideas in this paper but allows a simpler presentation.
The equilibrium condition reads
\begin{equation}\label{eq:stat}
	\int_{\Ome} \bfsig: \bfeps(\bfu^*) \dint \Ome  = 0, \quad  \forall \bfu^* \in U^*
\end{equation}
where
\begin{equation}
	U^*= \{ \bfu \in H^1(\Ome): \bfu = 0 \texton \Gam_u \}
\end{equation}

Kinematics and equilibrium  equations (\ref{eq:kin}-\ref{eq:stat}) must be 
complemented with the constitutive model.
We consider the formalism of generalized standard material introduced  in \cite{Halphen75,Germain83}.
The set of internal variables is denoted $\alpha$. It is a generic notation that  describes a set of scalar, vectorial or  tensorial variables.
The model is characterized by a free energy potential $\varphi(\bfeps, \alpha)$ and a 
dissipation potential $\psi(\dotalp, \alpha)$. 
We then introduce an  implicit time-discretization and use the energetic variational approach. 
Consider the displacement and internal variables $(\bfu_n, \alpha_n)$ known at some instant $t_n$. Finding the pair 
$(\bfu_{n+1}, \alpha_{n+1})$
at the next instant $t_{n+1} = t_n + \Delta t$ amounts to a 
minimization problem  
\begin{equation}
	(\bfu_{n+1}, \alpha_{n+1}) = \arg \min_{\substack{\bfu' \in U_n \\ \alpha' \in A_n}} F(\bfu', \alpha'; \bfu_n, \alpha_n, \dt)
\end{equation}
where $U_n$ is a short-hand notation  for $U(t_{n+1})$ and 
$A_n$ indicates the restriction on  $\alpha_{n+1}$.
The incremental potential $F$ involves the energy and dissipation potentials $\varphi$ and $\psi$.
Examples of incremental potentials will be given  in Section \ref{sec:mod}.

For simplicity, we shall consider time-independent material models.
In this case, the $F$ expression does not depend explicitely on $\bfu_n, \alpha_n$ and $\dt$. 
The extension to time-dependent models 
does not introduce difficulties.
Also, to lighten notation, we drop the $n+1$ indices.
The minimization problem is then 
\begin{equation}\label{eq:minau}
	(\bfu, \alpha) = \arg  \min_{\substack{\bfu' \in U_n \\ \alpha' \in A_n}} F(\bfu', \alpha') 
\end{equation}

We assume that the domains $U_n$ and $A_n$ are convex and  that $F$   is strictly convex with respect to the pair $(\bfu, \alpha)$ over $U_n \times A_n$. 
% These conditions ensures that the minimization yields an existing unique solution at each time-step. 
This defines a non-softening model.
The minimization is traditionally solved by a repeated sequence of two steps: the  computation of the internal variables (and stress) for a given displacement field, 
followed by the correction of the displacement field.  At iteration $m$, the two steps are given below 
\begin{align}
	\alpha^{m+1} & = \arg \min_{\alpha \in A_n}  F(\bfu^m, \alpha) \label{eq:un_min} \\
     \{\bfu^{m+1}\} & = \{\bfu^m  \}+  (K^m)^{-1} R^m, \quad %\text{\ correction\ }, 
      \bfu^{m+1} \in U_n \label{eq:deux_min}
%	\bfu^{m+1} & = \bfu^m - \left[  H^m \right]^{-1} \nabla_u F(\bfu^m, \alpha^{m+1}) 
%	=  \bfu^m - \left[  H^m \right]^{-1} R^m
\end{align}
The first step, \eqref{eq:un_min}, is purely local and may be carried out independently at each material point.
The second step, \eqref{eq:deux_min},  involves a linear solve
which updates the degrees of freedom \{\bfu\} associated to 
the field $\bfu$. The matrix $K$  
depends on the current internal variables $\alpha^{m+1}$ and 
the residual vector $R$ depends on the current stress $\bfsig^{m+1}$. 
The matrix can be the
algorithmic tangent operator \cite{Simo1997} or some approximation of it.
The linear solver may possibly be followed by a line search to further improve the solution.
Expressions of $K$ and $R$ are given  in  the Appendix for the models at stake in this paper.

\section{Softening variable and Lipschitz regularization} \label{sec:dam}
We now consider that the model has an extra scalar variable $d$ responsible for  softening. The optimization problem becomes
\begin{equation}
	(\bfu, \alpha, d) = \arg \min_{\substack{(\bfu', \alpha') \in U_n \times A_n  \\ d' \in D_n}}
	F(\bfu', \alpha', d')
\end{equation}
It is no longer convex, we can expect several local minima and also 
non-uniqueness to the global minimum (i.e. several solutions leading to the same 
global minimum).
Even though $F$ is not convex with respect to the triple $(\bfu, \alpha, d)$, we ask the optimization 
to be convex with respect to the couple $(\bfu, \alpha)$ for all  $d \in D_n$ and to be 
convex with respect to $d$ for all  $(\bfu, \alpha)
\in U_n \times A_n$.

Given a $d$ field over the domain $\Ome$, the Lipschitz constant associated 
to this field is the minimum $M$ value such that the following holds
\begin{equation}
	\abs{ d(\bfx) - d(\bfy)}  \leq M \, \dist(\bfx, \bfy), \quad \forall \bfx, \bfy \in \Ome 
\end{equation}
where $\dist(\bfx, \bfy)$ is the minimal length of the path inside $\Ome$ joining $\bfx$ and $\bfy$ (the distance is considered infinite if the two points cannot be connected inside $\Ome$). The $\dist$ function is a metric since it  satisfies for all $\bfx, \bfy, \bfz \in \Ome$:
\begin{align}
	& \dist(\bfx, \bfy) = 0 \Leftrightarrow  \bfx = \bfy \\
		& \dist(\bfx, \bfy)  = 	\dist(\bfy, \bfx) \\
   & \dist(\bfx, \bfy)  \leq  \dist(\bfx, \bfz) + \dist(\bfz, \bfy)
\end{align}
The value $M$ defined above is  denoted $\lip(d)$.
We define the regularization space for the damage field 
\begin{equation}
	\Lip = \{ d \in L^{\infty}(\Ome): \lip(d) \leq 1/l \}
\end{equation}
%\begin{equation}
%	d \in L \Leftrightarrow  \abs{ d(\bfx) - d(\bfy)}  \leq  \frac{\dist(\bfx, \bfy)}{l},  \forall \bfx, \bfy \in \Ome 
%\end{equation}
where $l$ is the regularizing length.
The set $L$ is convex.
We seek the solution as "one" of the global minima
\begin{equation}
		(\bfu, \alpha, d) = \arg \min_{\substack{(\bfu', \alpha') \in U_n \times A_n \\ d' \in D_n \cap \Lip}}
	F(\bfu', \alpha', d')	
%	(\bfu, \alpha, d) = \arg \min_{\substack{\bfu \in U_n \\ \alpha \in A_n \\ d \in D_n \cap L}}
%	F(\bfu, \alpha, d)
\end{equation}
We proceed by alternate minimization.
\begin{align}
	(\bfu^{k+1}, \alpha^{k+1}) & = \arg \min_{(\bfu, \alpha) \in U_n \times A_n} 	F(\bfu, \alpha, d^k) \label{eq:first_min} \\
	d^{k+1} & =   \arg \min_{d \in D_n \cap \Lip}  
    F(\bfu^{k+1}, \alpha^{k+1}, d) \label{eq:second_min}
\end{align}

\begin{figure}[h]
 	\centering
 	\def\svgwidth{0.75\textwidth} 
 	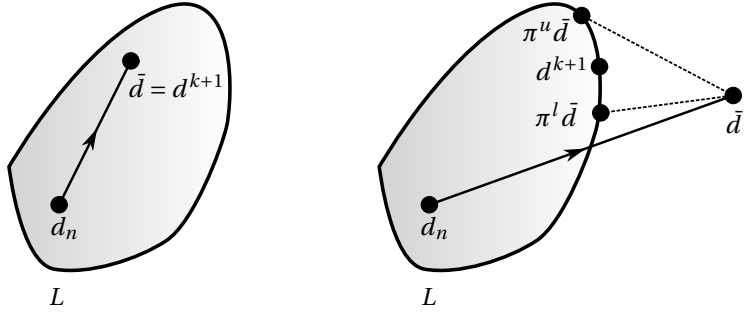  
 	\caption{A sketch of the local update $\od$ from the previous known damage field $d_n$. If the update satisfies the Lip constraint (left), 
 		we have directly the solution $d^{k+1}$. Otherwise (right), the local update needs to be projected back to the Lip constraint while minimizing the objective function $F$. The upper and lower projections give bounds to the solution.}
 	\label{fig:lip_proj}
 \end{figure}

For the  first minimization, the damage variable is frozen and the problem  is thus identical to problem \eqref{eq:minau}. 
It is a classical non-softening step.
The second minimization, \eqref{eq:second_min}, 
is less common. The objective function  is convex (and separable) as well as the constraint $D_n$.  The Lip constraint is non-local as it ties the damage variables between points. The optimization to find $d$
is thus potentially time-consuming when turning to a numerical implementation.
The good news is that the quest for $d$ may be decomposed into three steps reducing dramatically the cost of the optimization. The first step is  to create a trial $d$ field denoted $\od$ by ignoring the Lipschitz constraint and performing a decoupled minimization at each point 
\begin{equation}
	\od =   \arg \min_{d \in D_n}   F(\bfu^{k+1}, \alpha^{k+1}, d) 
\end{equation}
 If the trial damage $\od$ satisfies the Lipschitz constraint, it is the solution we are looking for as indicated in Figure \ref{fig:lip_proj} (left).
 If not, the optimal damage field will be different from $\od$.
 
\begin{figure}[h]
	\centering
	\includegraphics[width = 0.75\textwidth]{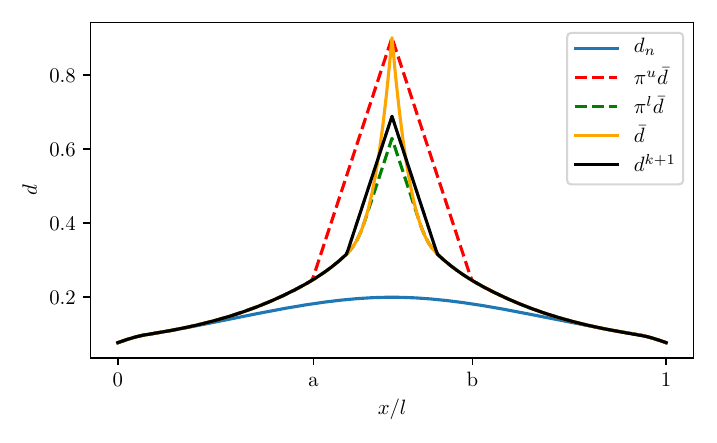}  
	\caption{A sketch of the bracketing capability of the upper and lower projections.  Outside the interval $[a,b]$, the projections are equal and 
	thus $d^{k+1} = \od$. Inside the interval  $[a,b]$, the projections bracket
     $\od$ and $d^{k+1}$.}
	\label{fig:bounds}
\end{figure}

We can find an upper bound of the domain 
 over which $d^{k+1}$ will differ from $\od$. 
We define two projections onto $L$, a lower projection $\pil$ and an 
upper projection $\piu$: 
\begin{align}
	\pil \od(\bfx) & = \min_{\bfy \in \Ome} \; (\od(\bfy) + \frac{1}{l} \dist(\bfx, \bfy)) \\
	\piu \od(\bfx) & = \max_{\bfy \in \Ome} \;  (\od(\bfy) - \frac{1}{l} \dist(\bfx, \bfy)) 
\end{align}
We prove in the appendix that these  projections satisfy the following inequality:
\begin{equation} \label{eq:bdbar}
	\dn \leq \pil \od \leq \od \leq  \piu \od \leq 1
\end{equation}
and provide bounds for the optimal damage
\begin{equation} \label{eq:bounds}
	\pil \od \leq d^{k+1} \leq  \piu \od 
\end{equation}
As an important consequence, the trial and  optimal solutions coincide wherever the bounds are equal
\begin{equation}
 \pil \od(\bfx) = \piu \od(\bfx)  \quad \Rightarrow \quad   d^{k+1}(\bfx) =  \od(\bfx)
\end{equation}
%Note that the Lipschitz point of view instead of bounded gradient is key to formulate these bounds.
A sketch of the projections is given in Figure \ref{fig:lip_proj} (right) and they are 
also illustrated  on the  one-dimensional  example, Figure 
\ref{fig:bounds}. 

\section{Elastic and plastic softening models} \label{sec:mod}
Consider the one-dimensional model of a bar of length $L$ and unit section  attached at its left side and pulled with an imposed displacement $\ud(t)$ at it's right end.  
The displacement at time $t$ must belong to the set
\begin{equation*}
	U(t) = \{ u \in C([0,L]): u(0) =0, \; u(L) = \ud(t)     \}
\end{equation*}
whereas the set of admissible displacement variations is given by
\begin{equation*}
	U^* = \{ u \in C([0,L]): u(0) =0, \; u(L) = 0    \}
\end{equation*}
We now detail several incremental potentials to be used in the 
simulation.

\subsection{Softening elasticity}
The elastic  potential reads
\begin{equation}
	\Fe(u) = \int_0^L  f_e(\eps(u)) \dint x, \quad \fe(\eps(u)) = \frac{1}{2} E \epsilon(u)^2
\end{equation}
where $\epsilon(u) = \frac{\mathrm{d} u}{\mathrm{d} x}$ and 
$E$ is the Young modulus. 
The elastic softening model, 
affects $E$ and adds a 
dissipation term.
For simplicity, we do not consider tension compression dissymmetry in the model (the bar will always be in tension in the simulation).
The softening elasticity potential is 
%\begin{equation}
%	\Fse(u, d) = \int_0^L \frac{1}{2}  (1-d)^2 E \epsilon(u)^2+  \Yc h(d) \dint x
%\end{equation}
\begin{equation} \label{eq:fse}
	\Fse(u, d) = \int_0^L \fse(\eps(u),d) \dint x, \quad 
	\fse(\eps(u),d) = (1-d)^2 \fe(\eps(u)) +  \Yc h(d)
\end{equation} 
where $\Yc$ is the critical energy release rate and the convex function  $h(d)$ defines the softening behavior.
Damage can only grow and cannot go beyond 1. 
The convex constraint set $D_n$ for damage is:
\begin{equation} 
	D_n = \{  d \in L^\infty([0,L]): d_n \leq  d \leq 1     \}  \label{eq:defD}
\end{equation}
The power 2 over the factor $(1-d)$ ensures convexity and 
a finite opening as $d \rightarrow 1$ \cite{Zghal2018}.
Note that the proper choice of the power is also an issue 
with gradient damage models for which a value
of 2  also ensures a finite opening  \cite{Lorentz2011a}.
We consider two choices for $h(d)$
\begin{align}
	h_1(d) & = 2 d + 3  d^2 \\
	h_2(d) & = \frac{2 d - d^2}{(1- d + \lam d^2)^2}
\end{align}
Both functions are convex (provided $\lam \leq 1/2$).
The condition $h'(0)=2$ ensures that when damage starts 
$ E \epsilon^2 / 2 = \Yc$.  
%The condition is $h''(0) = 3 h'(0)$
%ensures that the stress strain curve has a zero slope
%when $d=0$.
The second choice allows to mimic a linear cohesive-zone model 
\cite{ParrillaGomez2015} where $\lam$ is defined by 
\begin{equation}
	\lam = 2 \frac{ \Yc l}{\Gc}
\end{equation}
with $\Gc$ denoting the toughness (energy per section area needed to break the bar).
We shall see that the choice of $h_2(d)$ allows to choose 
independently $\Yc$, $\Gc$ and $l$.

\begin{figure}[h]
	\centering
	\includegraphics[width = 0.75\textwidth]{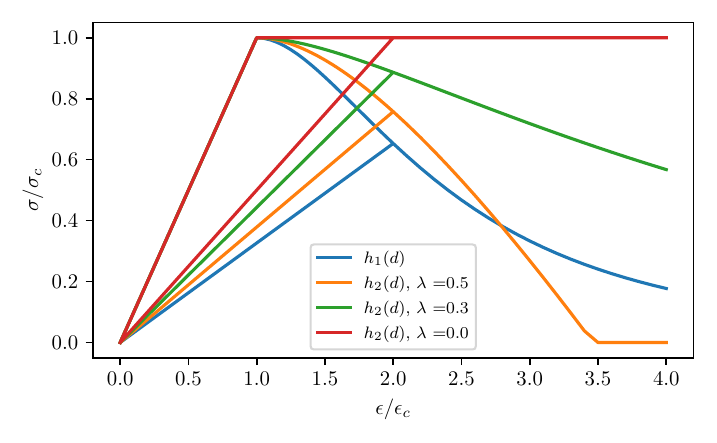}  
	\caption{Stress-strain relations for the softening elastic model ($\sigc = \sqrt{2 E \Yc}, \epsc = \sigc/E$). 
	The strain loading/unloading follows the peak values sequence: $\eps/\epsc = (0, 2, 0,4)$.}
	\label{fig:soft_elas}
\end{figure}

To get a better insight on the potential, we write the 
Karush-Kuhn-Tucker (KKT) conditions associated to the optimization.
We define the dual quantities to the strain and damage by taking 
the derivative of $\fse$. We get the stress 
$\sig$ and a variable  associated to damage denoted $\mu$ (damage criterion):
\begin{align}
	\sigma &= (1-d)^2 E \epsilon \\
	\mu & =  - (1-d) E \epsilon^2 +  \Yc h'(d)
\end{align}
where $h'(d)$ denotes the derivative of $h$ with respect to $d$.
The KKT conditions read 
\begin{align} 
	&\mu - \lam_1 + \lam_2  =  0  \label{eq:kktmu1}\\
	&\lam_1  \geq 0, \quad d-d_n \geq 0, \quad \lam_1(d-d_n) = 0 \label{eq:kktmu2} \\
	&\lam_2  \geq 0, \quad 1-d \geq 0, \quad \lam_2 (1-d) = 0 \label{eq:kktmu3}
\end{align}
where $\lam_1$ and $\lam_2$ are Lagrange multipliers associated 
to the constraints. We note that $\mu=0$ when damage is growing (and stays below 1), and $\mu \geq 0$ when damage 
does not grow (and is different from 1). Figure \ref{fig:soft_elas} shows the stress-strain relation for a strain loading-unloading history.

The minimization of 
$F_{se}$ under the constraint $(u,d) \in U_n \times D_n$ is 
not a convex problem. But, the minimization of $F_{se}$ with 
respect to $u \in U_n$, for $d \in D_n$ fixed  is a convex problem.
The minimization of $F_{se}$ with 
respect to $d \in D_n$ for $u \in U_n$ fixed is also a convex problem.
Finally, note that in the presentation above, 
we did not care in presenting separately the free energy  and 
dissipation potentials. This is not essential since we are not studying the temperature evolution. Moreover, for a given stress-strain relation this choice is in general non-unique.

\subsection{Softening elasticity with hardening plasticity}
The introduction of softening in elasto-plastic models is a complex
topic. The goal here is not to find the appropriate model for a given situation
but rather to discuss how the Lip-field approach is behaving 
in the  presence of plastic internal variables. We consider 
a  basic von Mises plasticity model with elastic  softening and, in the next section, a softening plasticity model with preserved elasticity.
The incremental potential for a 
von Mises isotropic hardening plasticity model is 
%\begin{equation}
%	\Fp(u, \ep, p) = \int_0^L  \frac{1}{2} E (\epsilon(u) - \epsilon_p)^2 +
%	\sigy ( p + \frac{k}{2} p^2)  \dint x \label{eq:Fp}
%\end{equation}
\begin{equation}
	\Fp(u, \ep, p) = \int_0^L  \fp(\eps(u), \ep, p) \dint x \label{eq:Fp}, \quad 
	\fp(\eps(u), \ep, p) = \frac{1}{2} E (\epsilon(u) - \epsilon_p)^2 +
	\sigy ( p + \frac{k}{2} p^2) 
\end{equation}
where $\sigy$ is the yield stress and $k$ the isotropic hardening 
parameter.
Regarding the constraints, plasticity internal variables 
must belong to the following convex set 
\begin{equation} \label{eq:defAn}
	A_n = \{  (\epsilon_p, p) \in (L^\infty([0,L]))^2:  p - p_n \geq \abs{\epsilon_{p}-\epsilon_{pn}}      \}
\end{equation}
The minimization of $\Fsp$ with respect to $(u,(\ep,p)) \in U_n \times A_n$ 
is a convex problem. 
We introduce damage  with a multiplicative approach, also called effective stress approach:
\begin{align}
	\Fsep(u, \ep, p, d) &  =  
	\int_0^L \fsep(\eps(u), \ep, p, d)  \dint x, \\
	\fsep(\eps(u), \ep, p, d) & = (1-d)^2  \fp(\eps(u), \ep, p) + \Yc h(d)
\end{align}
The damage must belong to the set $D_n$ defined in 
\eqref{eq:defD}.

\begin{figure}[h]
	\centering
	\includegraphics[width = 0.75\textwidth]{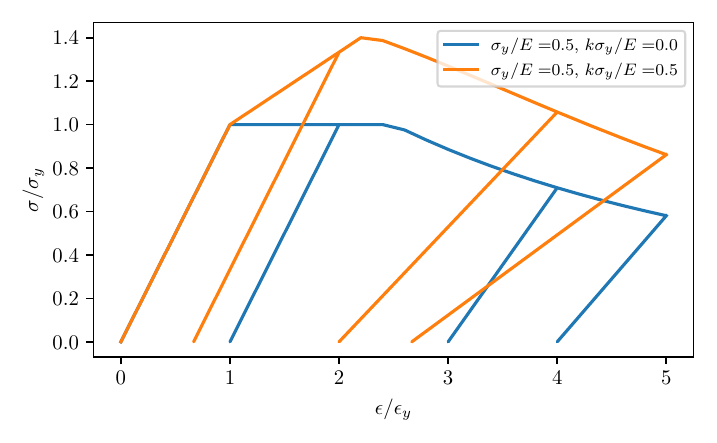}  
	\caption{Stress-strain relations for the softening elasticity model with hardening plasticity with the choice $\epsc/\epsy = 2$ where $\epsy = \sigy/E$. 
		The strain loading/unloading follows the peak value
		sequence $\eps/\epsy = (0, 2, e_1, 4, e_2, 5, e_3)$ where $e_1$,  $e_2$, $e_3$ indicate the strain needed to reach full unloading ($\sig = 0$).}
	\label{fig:soft_comb}
\end{figure}

Taking the derivative of  $\fsep$ 
with respect to the plastic strain, cumulative plasticity and damage, gives the stress, current yield stress and the damage 
criterion, respectively
\begin{align}
	\sigma &= (1-d)^2 E (\epsilon - \epsilon_p) \\
	R & =  \sigy (1-d)^2 ( 1 + k p) \\
	\mu & = - (1-d)  E (\epsilon(u) - \epsilon_p)^2   -  2 (1-d) \sigy (p + k p^2/2 )    +  \Yc h'(d) \label{eq:muelas}
\end{align}
They are involved in the KKT conditions 
\begin{align}
	& \sig + \lam_\mathrm{p}  s(\epsp-\epspn) = 0 \\
	& R - \lam_\mathrm{p} = 0 \\
	& \lam_\mathrm{p}  \geq 0, \quad p - p_n - \abs{\epsp-\epspn} \geq 0, 
	\quad \lam_\mathrm{p}  (p - p_n - \abs{\epsp-\epspn}) = 0
\end{align}
where $s$ is the multi-valued signed function 
($s(x) = -1$  if  $x<0$, $s(x) = +1$ if $x > 0$ and $ s(0) \in [-1,1]$).
We observe that the effective stress $\sig/(1-d)^2$ 
and variables $\epsp$ and  $p$ may be computed from the strain independently of the damage variable.  Stress-strain relations for  a strain loading-unloading history are shown in Figure \ref{fig:soft_comb}. The effect of damage may 
be observed in the unloading phase.

\subsection{Softening plasticity}
Finally, we consider an elasto-plastic model in which the softening affects only the yield stress leaving  elasticity  unchanged:
\begin{align*} 
	& \Fsp(u, \ep, p,d)  = \int_0^L  \fsp(\eps(u),\ep,p,d) \dint x, \\
	& \fsp(\eps(u),\ep,p,d) =  \frac{1}{2} E (\epsilon(u) - \epsilon_p)^2 +
	 (1-d)^2 \sigy  (p + k p^2/2) + \sigy g(d)
	\label{eq:Fsp}
\end{align*}
where $g(d)$ describes softening and is chosen 
as $g(d) = d^2$.
\begin{figure}[h]
	\centering
	\includegraphics[width = 0.75\textwidth]{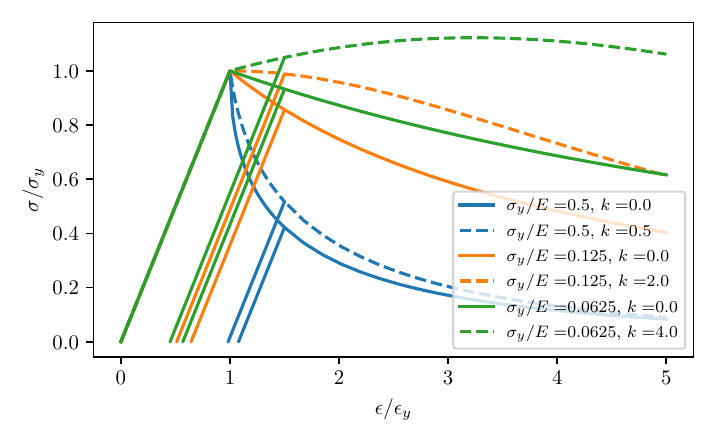}  
	\caption{Stress-strain relations for the softening plastic model ($\epsy=\sigy/E)$. The strain loading/unloading follows the peak value sequence  $\eps/\epsy = (0, 1.5, e, 5)$ where $e$ indicates the value needed to reach $\sig = 0$.}
	\label{fig:soft_plas}
\end{figure}

Taking the derivative of $\fsp$, we get
\begin{align}
	\sigma &= E (\epsilon - \epsilon_p) \\
	R & =  \sigy (1-d)^2  (1 + kp) \\
	\mu & = - 2 \sigy (1-d) (p + k p^2/2) + \sigy g'(d) \label{eq:muvoid}
\end{align}
and the associated  KKT conditions 
\begin{align}
	& \sig + \lam_\mathrm{p}  s(\epsp-\epspn) = 0 \\
	& R - \lam_\mathrm{p}  = 0 \\
	& \lam_\mathrm{p}  \geq 0, \quad p - p_n - \abs{\epsp-\epspn} \geq 0, 
	\quad \lam_\mathrm{p}  (p - p_n - \abs{\epsp-\epspn}) = 0
\end{align}
%& 	f = \abs{\sigma} - R \leq 0, \quad p - \pn \geq 0, \quad   f  (p - \pn) = 0 \label{eq:f}\\
%&	Y \leq 0, \quad d-\dn  \geq 0, \quad   Y (d-\dn) = 0 \label{eq:Y}

We have omitted above the  equations associated to $d$. They are identical to the ones presented in 
(\ref{eq:kktmu1}-\ref{eq:kktmu3}). 
\textcolor{black}{The growth of $d$ is now linked to cumulative plastic strain and no longer to the elastic strain (as indicated by the difference between  \eqref{eq:muelas} and \eqref{eq:muvoid}).}
%The $d$ variable may be interpreted as 
%a void volume fraction quantity used in ductile failure %models \cite{Tvergaard1984,Becker1988}.
Stress-strain relations for  a strain loading-unloading history are shown in Figure \ref{fig:soft_plas}. The elasticity is not affected as can be observed
from the unloading phase.

 \begin{figure}[h]
 	\centering
 	\def\svgwidth{0.75\textwidth} 
 	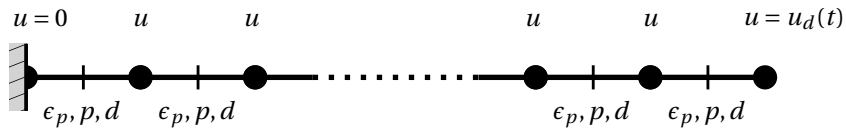  
 	\caption{A one-dimensionbal bar discretized with equal-sized finite elements. Displacements are stored at the node whereas internal variables are stored at the element integration point (depicted with a vertical bar).}
 	\label{fig:mesh_dof_1d}
 \end{figure}

\section{Lipschitz regularization over a bar} \label{sec:fem}
The bar is discretized with $N$ finite elements of equal size $h=L/N$.
The displacement is linear over each element between nodal values. The internal variables are stored at each element centroid (element integration point) as indicated in Figure \ref{fig:mesh_dof_1d}.
The $i$ index is used either for the element numbering
( $d_i, i = 1, \ldots, N$  denote the damage at the centroid of each element) or for the node numbering ($u_i, i = 1, \ldots, N+1$ are the nodal displacements).
The Lipschitz constraint implies the following inequalities defining the set $L$
\begin{align*}
	d_i - d_{i+1} - h/l & \leq 0, \quad  i= 1, \ldots, N-1 \\
	d_i - d_{i-1} - h/l & \leq 0, \quad i= 2, \ldots, N 
\end{align*}
The optimization problem reads for the softening elasticity problem
\begin{equation}
	\min_{\substack{\bfu \in U_n \\ d \in D_n \cap L}}
	\sum_{i=1}^{N} h \fse( \frac{u_{i+1} - u_i}{h},  d_i) 
\end{equation}
and 
\begin{equation}
	 \min_{\substack{(\bfu, (\epsp, p)) \in U_n \times A_n \\ d \in D_n \cap L}}
	\sum_{i=1}^{N} h \fx( \frac{u_{i+1} - u_i}{h}, \epspi, p_i, d_i) 
\end{equation}
for the plasticity models 
($\mathrm{x}$ stands for  $\mathrm{sep}$ or $\mathrm{sp}$).
The set $U_n$ enforces that $u_0=0$ and $u_{N+1} = \ud(t_{n+1})$.
The set $A_n$ is related to \eqref{eq:defAn} and the set $D_n$ enforces the fact that $d_i$ must be above its previous time-step value and below 1.
For all examples treated, the  softening elasticity function is chosen as $h_2$.
%and the plasticity softening function $g(d)$ is chosen as $d^2$.

\begin{figure}[h]
	\centering
	\begin{tabular}{cc}
		\includegraphics[width = 0.49\textwidth]{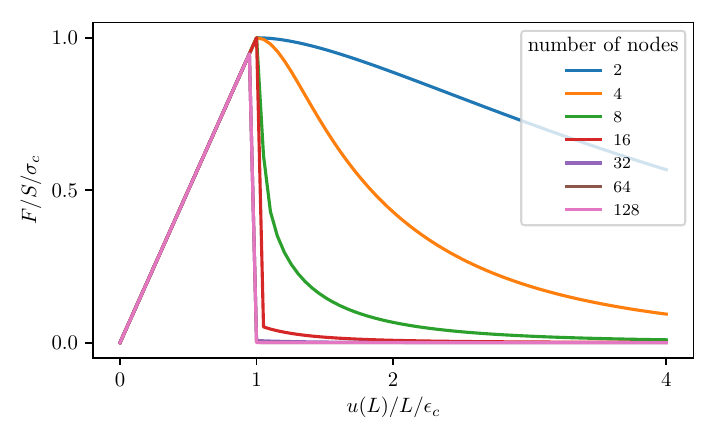}  &
		\includegraphics[width = 0.49\textwidth]{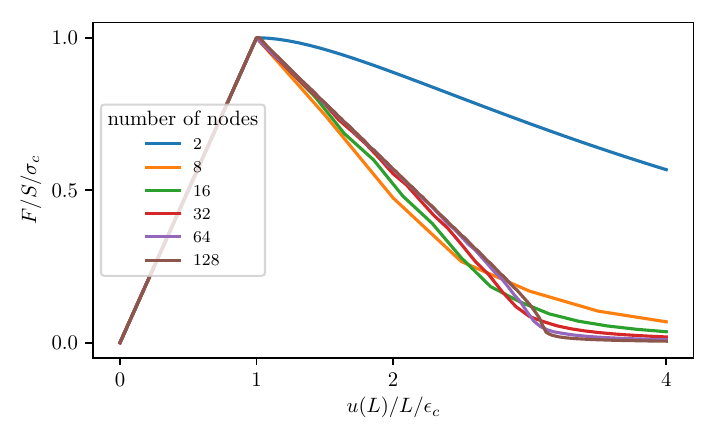} \\
			\includegraphics[width = 0.49\textwidth]{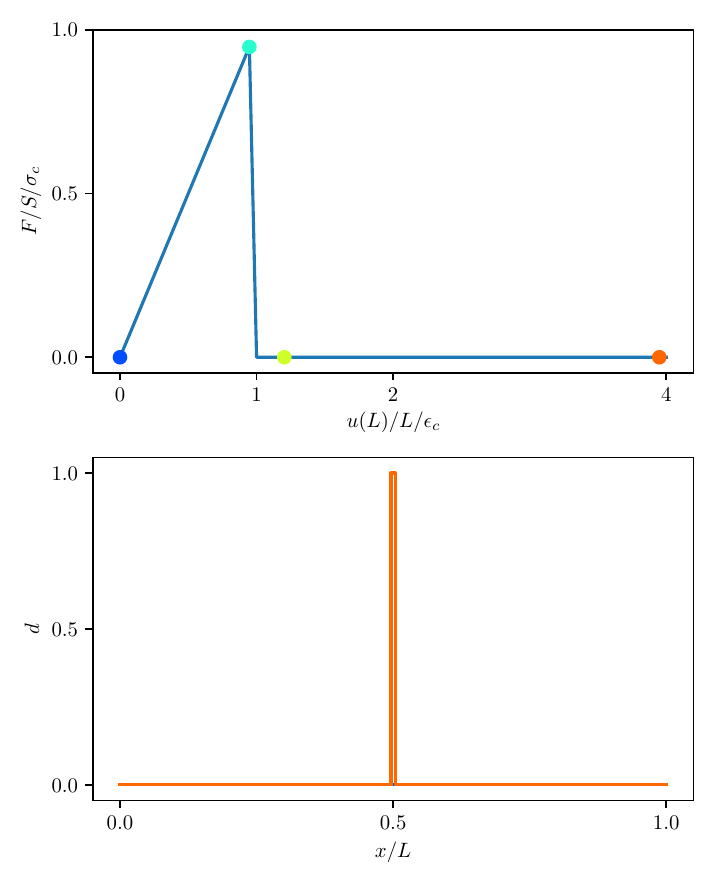}  &
		\includegraphics[width = 0.49\textwidth]{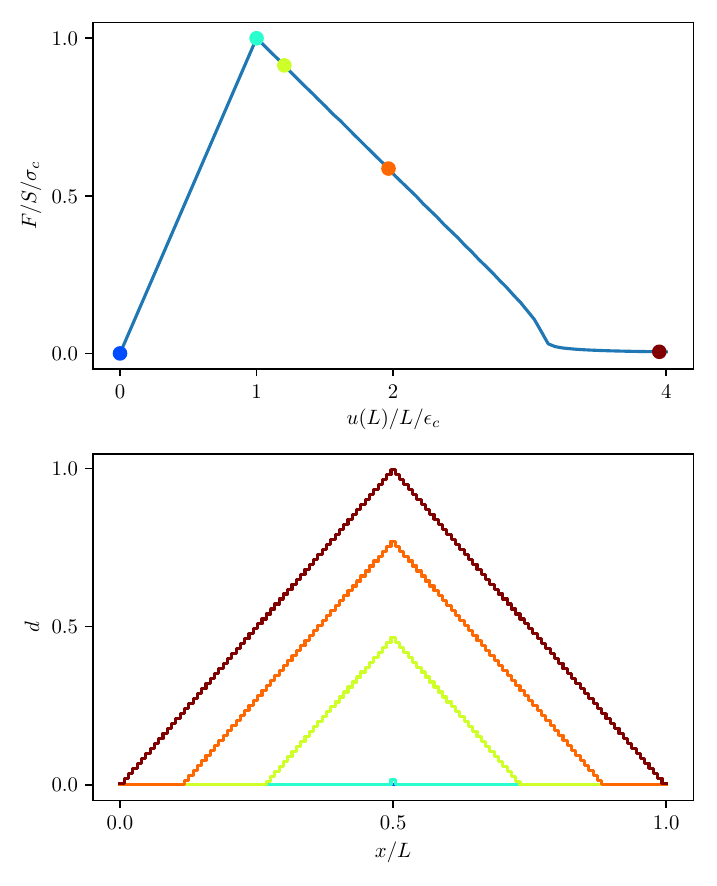}
	\end{tabular}
	\caption{Response for the bar with the elastic softening behavior and imposed displacement:  non-regularized model (left) and Lip-field model (right). The top row gives the stress-strain average response and the bottom one the damage profile 
		at the color dots given in the middle row.  
	The mesh used for the middle row is the most refined of the top row. 
	  Parameters are:
		$L$=1,  $l$ = 0.5, $E$ = 1, $\Yc$ = 1, $\lam$ = 0.3, $\Gc$  = 10/3. }
	\label{fig:elas_soft_l05}
\end{figure}

The optimization problems above are solved by alternate minimization as explained in 
Section \ref{sec:dam}. The minimization with respect to the damage variable is performed using the scipy python library.
More precisely, the  scipy.optimize.minimize 
function is used. It is described in  
\cite{2020SciPy-NMeth}. It offers 
an interface to the  SLSQP 
(Sequential Least Square Quadratic Programming) routine created  
by Dieter Kraft in  \cite{kraft1988software}.

Because of its homogeneity, the optimation problem at stake has no 
unique global minimizer and many local minimizers. 
Aware of this situation, we trigger localization by introducing a 
slightly higher damage on top of $d_n$ in the middle of the bar at each step. 
This is just  an initial guess for the alternate minimization.
This strategy does not require to alter the stiffness or initial damage 
on the bar. 

\begin{figure}[h]
	%\centering
	\begin{tabular}{cc}
		\includegraphics[width = 0.49\textwidth]{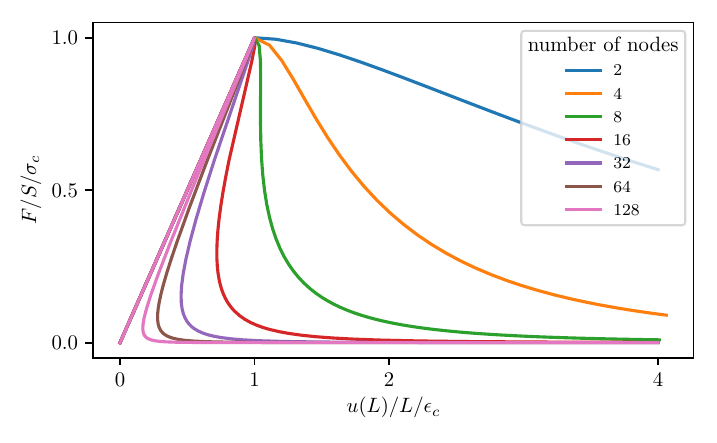}  &
		\includegraphics[width = 0.49\textwidth]{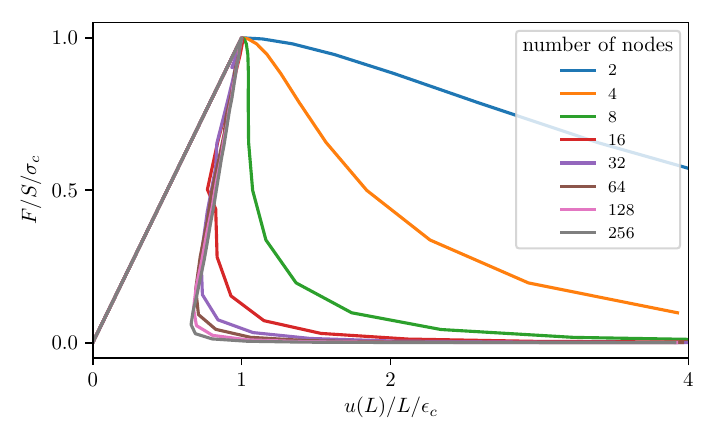} \\
				\includegraphics[width = 0.49\textwidth]{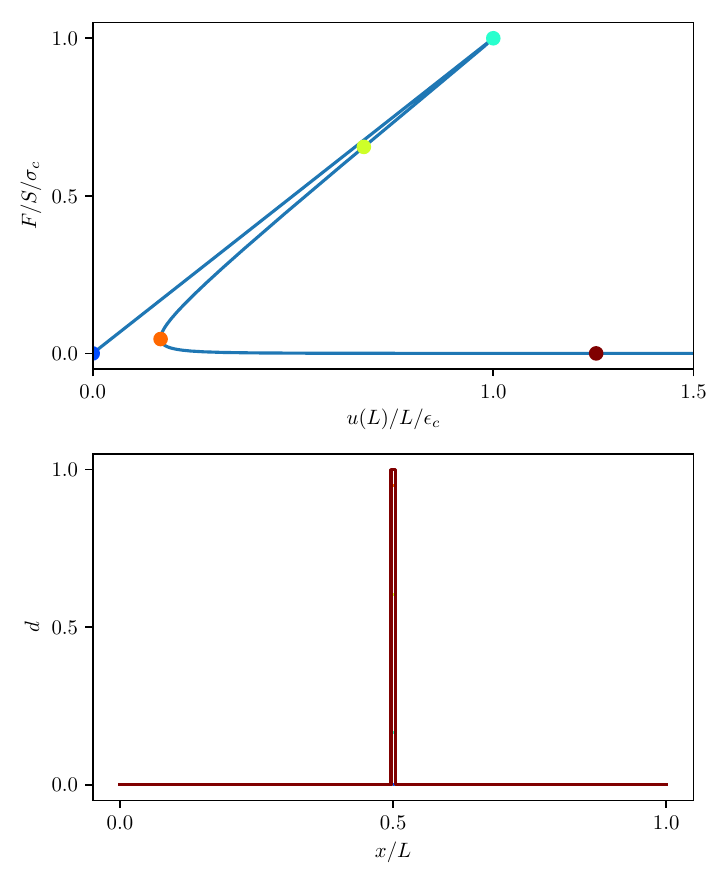}  &
		\includegraphics[width = 0.49\textwidth]{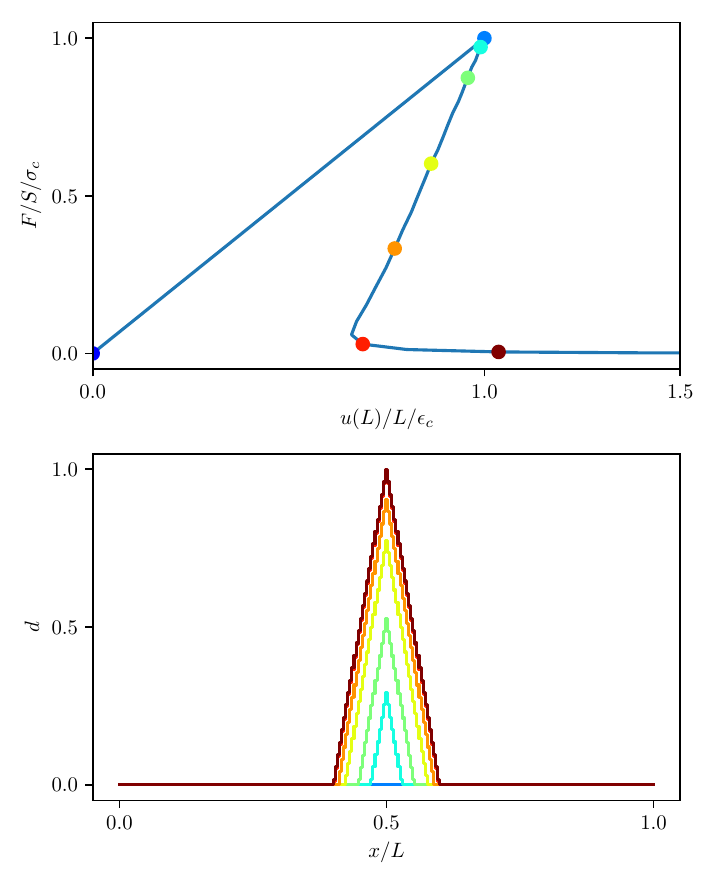}
	\end{tabular}
	\caption{Response for the bar with the elastic softening behavior and snap-back control:  non-regularized model (left) and Lip-field model  (right), convergence of the stress-average strain response (top) and damage evolution (bottom) corresponding the color dots in the middle figures.
	The middle figures reproduce the curve of the top figures for the most refined grid.  
	Parameters are:
		$L$=1,  $l$ = 0.1, $E$ = 1, $\Yc$ = 1, $\lam$ = 0.3, $\Gc$ = 2/3.}
	\label{fig:elas_soft_snap}
\end{figure}

For the elastic softening model, Figure \ref{fig:elas_soft_l05} 
gives the bar response for the 
non-regularized and Lip-field models.
The regularizing effect of the latter is clear.  
Note that 
the case of two nodes for the discretization 
(a single element) gives the 
homogeneous bar response.
No snap-back appears in this example because the regularization length is pretty large.
A smaller length and toughness are used for Figure \ref{fig:elas_soft_snap}
and a snap-back may be observed. To handle the snap-back in the simulation, the displacement at the end of the bar is no longer imposed but controlled by limiting the strain increment over the bar at each time step. More evolved control could be used  \cite{Crisfield1981,DeBorst1987}. 
Figure \ref{fig:elas_soft_snap} shows for the 
non-regularized case (left column) 
a convergence towards 
a solution without any dissipation (zero area under the converged curve). On the contrary, the converged solution, (right column) indicates a non-zero dissipation (whose dimensional value is close to  the toughness $\Gc$).

%\begin{figure}[h]
%	\centering
%	\begin{tabular}{cc}
%	\end{tabular}
%	\caption{Corresponding damage history to the previous figure. The color code of the dots in the upper Figure gives the corresponding damage field in the bottom figure.}
%	\label{fig:elas_soft_snap_d}
%\end{figure}

%\begin{figure}[h]
%	\centering
%	%\includegraphics[width = 0.75\textwidth]{noarc.pdf}  
%	\caption{Stress-average strain response for the bar with the elastic softening behavior and snap-back possibility  and an evolving number
%		of elements :  non-regularized (left) and Lip-field model response (right).}
%	\label{fig:noarc}
%\end{figure}

\begin{figure}[h]
	\centering
	\includegraphics[width = 0.55\textwidth]{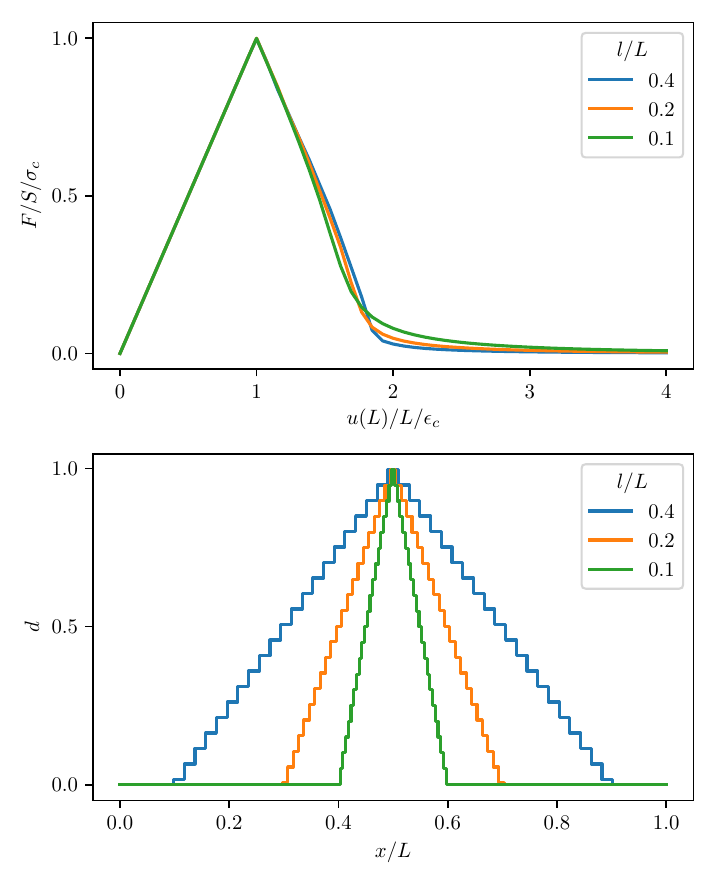}  
	\caption{Lip-field model response with the elastic softening behavior (choice $h_2$) for several regularization length values. Stress-average strain response (above) and damage profile (below).  Parameters are: $L$ = 1., $l$ = (0.4, 0.2, 0.1), $E$ = 1.,  $\Yc$ = 1.,    $\lam$ = (0.4,0.2,0.1), $\Gc$ = 2.
	The number of elements is 51, 101 and 201 so that we have  $h/l \sim 20$.}
	\label{fig:elas_soft_coh}
\end{figure}

Still for the elastic softening model, we  show, Figure \ref{fig:elas_soft_coh}, that the stress-strain response does not vary much with the regularization length (demonstrating the 
nice capability of the $h_2$ softening function). 
The response are obtained for a non-varying toughness, $\Gc$, and
 strength $\Yc$. Only the regularization length is varied
 (and the mesh size is adjusted to always have the same number
 of elements over $l$). We observe a dimensional area 
 under the curve close to $\Gc$. 

\begin{figure}[h]
%	\centering
\begin{tabular}{cc}
	\includegraphics[width = 0.49\textwidth]{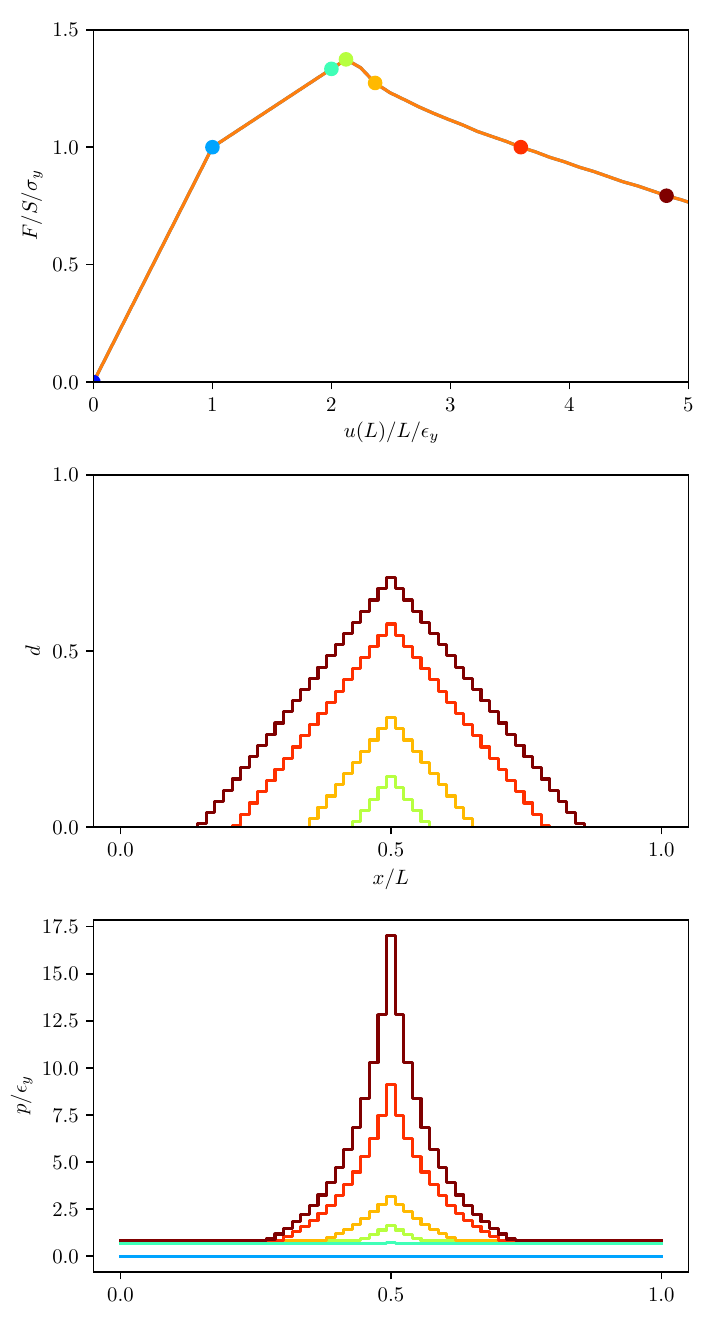}  &
	\includegraphics[width = 0.49\textwidth]{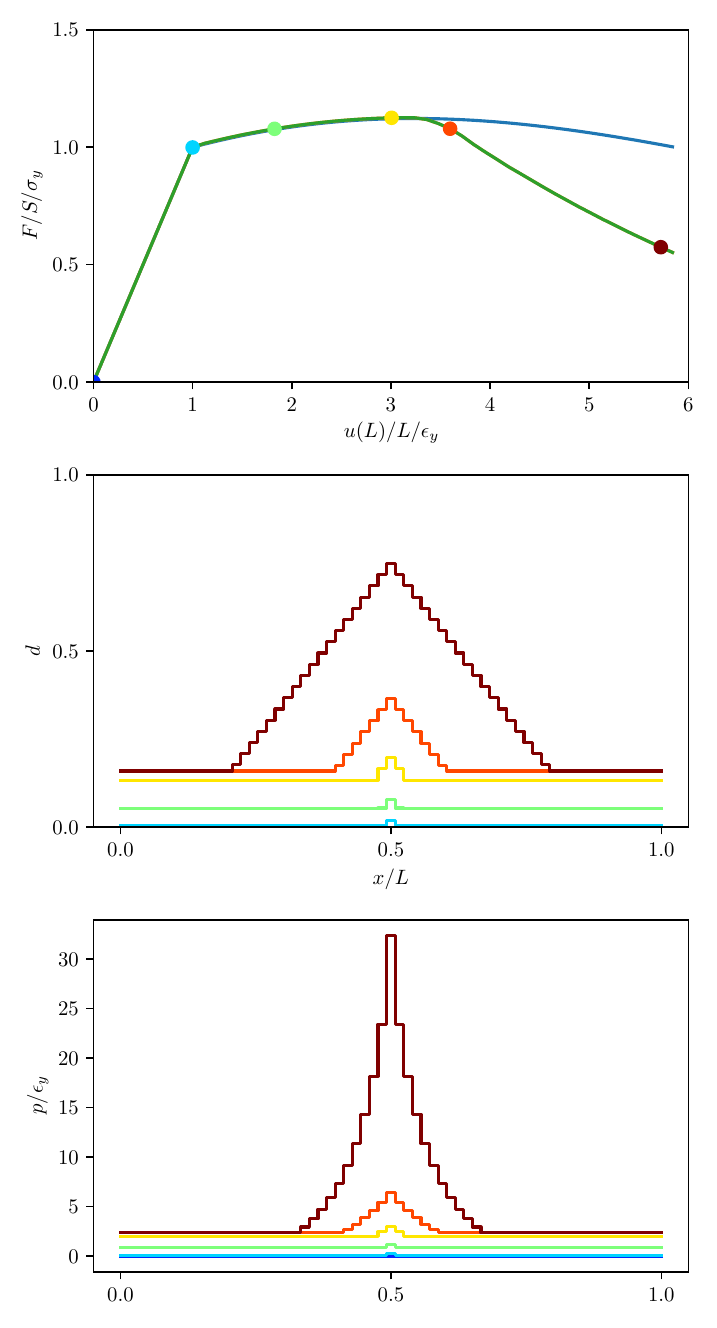}
\end{tabular}
	\caption{Lip-field model response for the softening elasticity model with plasticity hardening (left column) and softening plasticity (right column): stress-strain average curves (top row), damage field (middle row) and 
		cumulative  plasticity (bottom row). For both models,  $L=1$, $l=0.5$ and the mesh size
		is $h=L/64$. Model 
		specific parameters are:  
		$E = 2$, $\sigy$ = 1, $\Yc$ = 1, k=1, $\lam$ = 1/3
		for the hardening plasticity (corresponding to the model the orange curve in Figure \ref{fig:soft_comb})
		and 
		E = 1., $\sigy$ = 1/16, k = 4 for the softening plasticity model
		(corresponding to the model with the dashed green curve in Figure \ref{fig:soft_plas}).}
	\label{fig:plas}
\end{figure}

Regarding the softening elastic model with hardening plasticity, the response is given 
in Figure \ref{fig:plas} (left column). We observe that plasticity proceeds first in an homogeneous fashion, without any damage.
During this homogeneous phase, the stress-average strain curve is rising. As this curve reaches its peak, localization of the 
damage and plasticity start to occur. 
For the softening plasticity model, the response is  given 
in Figure \ref{fig:plas} (right column). 
We observe that plasticity and damage proceed first in an homogeneous manner due to  the initial hardening effect of plasticity, then concentrates. The  departure from the homogeneous response also occurs 
when the stress-average strain reaches its limit point.

\begin{figure}[h]
%	\centering
		\begin{tabular}{cc}
	\includegraphics[width = 0.49\textwidth]{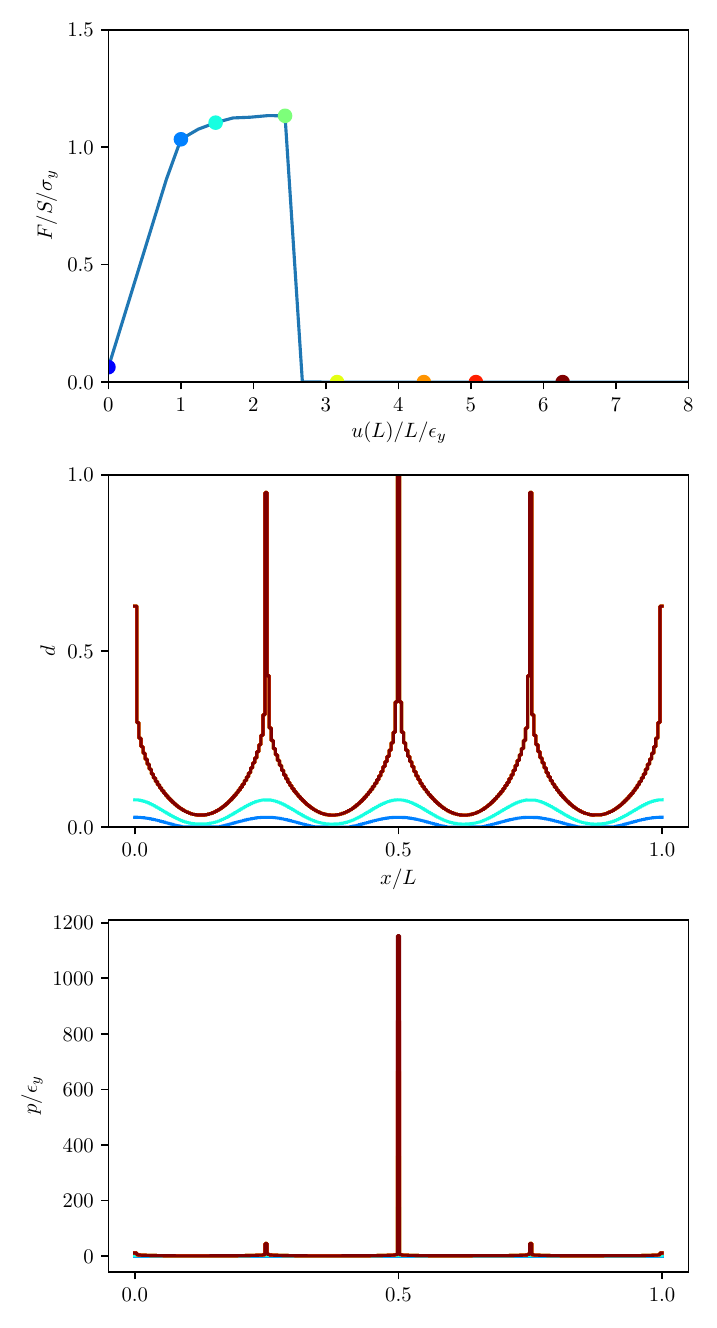}  
	&
	\includegraphics[width = 0.49\textwidth]{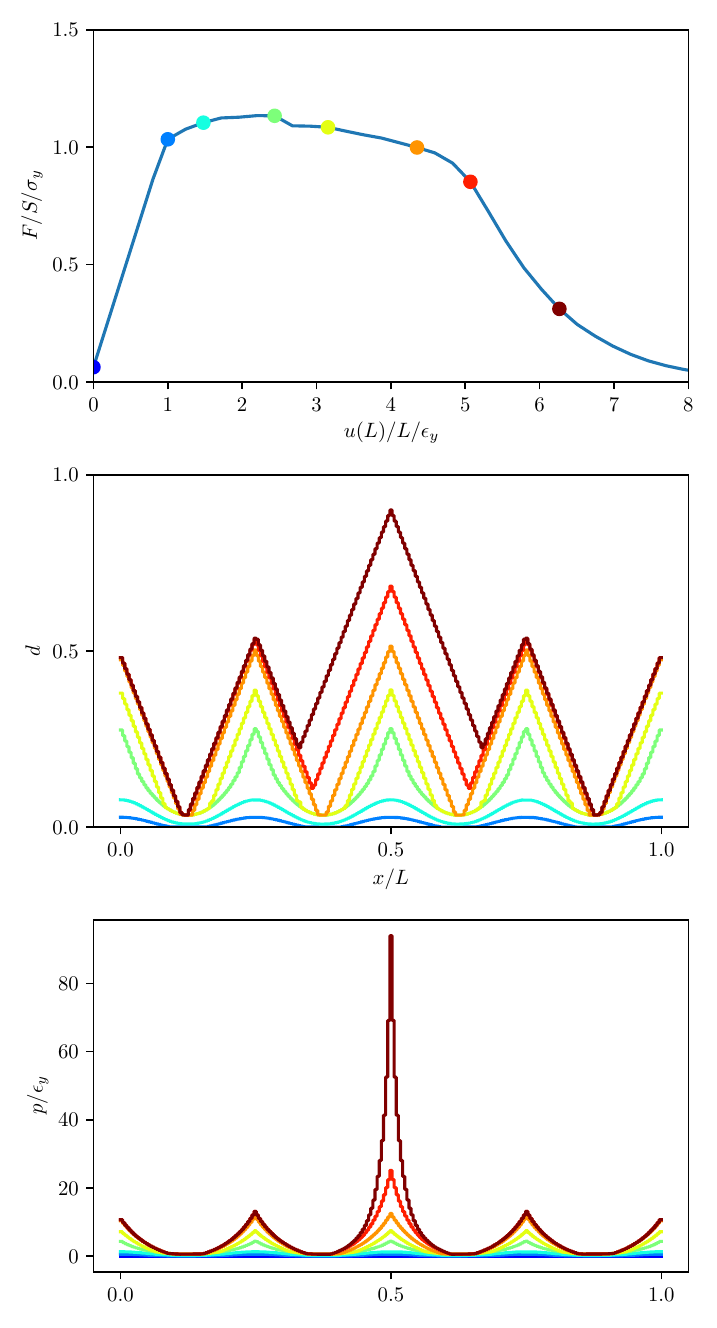}  
	\end{tabular}
	\caption{Response for a softening plasticity model with a volumic oscillatory force. No regularization, $l=0$ (left) and 
		Lip regularization (right). 
		Stress-average strain results (top), damage (middle) and cumulative plasticity profile (bottom).  Parameters are: $L$ = 1., $l$ = 0.25, $h$ = 1/255, $E$ = 1., $\sigy$ = 1/16,
		$k$ = 4. }
	\label{fig:plas_soft_extforce}
\end{figure}

We provide a last example with a non-uniform loading caused by  a volumic force $f(x/L) =  0.1*\sin(8*\pi*x/L)$. 
This loading is added to the imposed displacement. 
The results are depicted in Figure  \ref{fig:plas_soft_extforce}.
The model parameters are the same as for Figure \ref{fig:plas} (right column) except that $l=0.25$
and the number of nodes is 256. The damage develops in a complex pattern. 
The right-column gives the results for the non-regularized case
(local approach $l=0$). We observe the Lip-field results (right column) are the same as the local approach as long as the stress-average strain curve did not reach the peak point. This demonstrates the fact that the Lip-field approach preserves the local solution when it is not localizing. This is quite different from the other regularizing approaches (non-local, damage gradient, phase-field) that will modify the local solution even
prior to localization.
\textcolor{black}{It may also be seen  from Figure \ref{fig:plas_soft_extforce} that the normal derivative of the damage is not zero at the boundary as it would have been the case for damage gradient based models. The Lipschitz constraint does not enforce a given value for the normal derivative of the damage on the boundary  but just bounds the value.}
%any specific boundary conditions.}

%\begin{figure}[h]
%	\centering
%	\includegraphics[width = 0.75\textwidth]{plas_soft_extforce.pdf}  
%	\caption{Lip-field response for a softening plasticity model. Same parameters and loading as for Figure \ref{fig:plas_soft_extforce}  but without regularization ($l=0$).}
%	\label{fig:plas_soft_extforce_0}
%\end{figure}

\section{Conclusion and future works} \label{sec:conc}

A new regularization approach has been introduced to 
alleviate spurious localization with softening material models. Both elastic and plastic 
softening models have been considered.
The regularization enforces a Lipschitz condition on the field responsible 
for the softening.  In doing so,  a length is introduced in the model.

%(field simply called damage in this paper). 

Compared to the gradient damage or phase-field approaches, the Lip-field approach 
introduces an extra constraint but does not affect the expression of the incremental potential (the objective function in the minimization is not modified).
In other words, the energy does not depend on the damage gradient but solely  on the damage. Its expression is the one of the local model. As a consequence, the Lip-field does not introduce an extra partial differential equation with its questionable boundary conditions.
\textcolor{black}{From a mathematical point of view, the Lipschitz constraint may be interpreted as searching  the damage field among subsolutions (in the viscosity sense) of the eikonal equation \cite{Crandall83,Crandall2001} or as viscosity solutions to an eikonal inequality.}

Compared to the non-local integral approach, no widening of the fully damage 
zone is observed and the implementation only requires an element to 
be related to its neighbors.
Finally, compared to the Thick Level Set approach, 
the level set know-how is no longer needed.

Future works will be dedicated to the extension of the results to two- and 
three-dimensional problems. For this extension, the bounds demonstrated 
in this paper  (valid for any dimensions) 
will allow one to predict a priori the zones over 
which damage is affected  by the Lipschitz constraint. 
This will reduce the cost of the damage optimization.

Finally, the paper did not discuss the possibility of allowing displacement jumps during the simulation. 
This issue is important when dealing with fragmentation or crack under large deformation and needs to be studied in the future for the Lip-field approach.

\section*{Acknowledgements}
The authors wish to thank J. Dolbow,  B. Le,  B. Masseron, G. Rastiello, J. Rethore,  A. Salzman,   I. Stefanou, A. Stershic
and C. Stolz,
for their advice on ways to improve the manuscript.

\clearpage
\appendix
\section{Proof of the bounds}
Consider a  field $d$ defined over $\Ome$.
To this field, we associate two other fields denoted, $\piu d$ and $\pil d$,  called upper and lower projections, respectively.
\begin{align}
	& \piu: L^\infty(\Ome) \mapsto \Lip: d(\bfx) \mapsto  \piu d(\bfx)   = \max_{\bfy \in \Ome} \;  (d(\bfy) - \frac{1}{l} \dist(\bfx, \bfy)), \label{eq:projs} \\
	& \pil: L^\infty(\Ome) \mapsto \Lip: d(\bfx) \mapsto  \pil d(\bfx)  = \min_{\bfy \in \Ome} \; (d(\bfy) + \frac{1}{l} \dist(\bfx, \bfy)) 
\end{align}
They satisfy the following properties:
\begin{enumerate}
\item[(a)] $\pil d \in \Lip$,  $\piu d \in \Lip $
\item[(b)] $d \in L \Rightarrow \pil d = \piu d = d$
\item[(c)] $\pil d \leq d \leq \piu d$
\item[(d)] $ d_1 \leq d_2 \Rightarrow  \pil d_1 \leq \pil d_2, \; 
\piu d_1 \leq \piu d_2$ (monotonicity)
\end{enumerate}
To prove that the mapping falls indeed in $\Lip$ (property (a)), 
we take the difference 
\begin{align}
	\piu d(\bfx^1)  - \piu d(\bfx^2) = \max_{\bfy \in \Ome} \;  (d(\bfy) - \frac{1}{l} \dist(\bfx^1, \bfy))   - \max_{\bfy \in \Ome} \;  (d(\bfy) - \frac{1}{l} \dist(\bfx^2, \bfy)) 
\end{align}
Using the triangular inequality written as 
\begin{equation*}
	\dist(\bfx^1, \bfy) \leq \dist(\bfx^1, \bfx^2) + \dist(\bfx^2, \bfy)
\end{equation*}
we get 
\begin{equation*}
	 \piu d(\bfx^1) - \piu d(\bfx^2) \geq - \frac{1}{l} \dist(\bfx^1, \bfx^2)	
\end{equation*}
and using  the triangular inequality as 
\begin{equation*}
	\dist(\bfx^2, \bfy) \leq \dist(\bfx^2, \bfx^1) + \dist(\bfx^1, \bfy)
\end{equation*}
we get 
\begin{equation*}
 \piu d(\bfx^1) - \piu d(\bfx^2) \leq  \frac{1}{l} \dist(\bfx^1, \bfx^2)
\end{equation*}
%showing that indeed $\piu d \in L$. 
The proof is similar for $\pil d$. 
Property (b) is obtained from the definition of $d \in \Lip$:
\begin{equation}\label{key}
	d(\bfy)  -\frac{1}{l}  \dist(\bfx, \bfy) \leq   d(\bfx)  \leq 
	d(\bfy)  +\frac{1}{l}  \dist(\bfx, \bfy)
\end{equation}
Computing  the minimum and maximum  over $\bfy$ for  the upper and lower bounds, respectively yields the property.
Property (c) is obtained directly from the definition of the projections by testing $\bfy$ as $\bfx$ in the max and min. Similarly, 
property (d) stems from the  definition of the projections. 

We now discuss the consequences of the properties.
Properties (a) and (b) indicate that the projections are 
idempotent (applied twice yield the same result as applied once).
Property (c) indicates that at any point where the projections are equal, they are equal to $d$. 
Combining properties (c) and (d)  and using the fact that $\dn \in  \Lip$ (as well as noting that any uniform function is Lipschitz)
yield the result \eqref{eq:bdbar}.

We are now ready to prove the bounds result \eqref{eq:bounds}. 
Consider a damage field $d^*\in \Dn \cap \Lip$ 
lying ouside the bounds, 
 we associate to this field another field by clipping it to the bounds
 \begin{equation}
 	d^{**}(\bfx)  = \max (   \pil \od(\bfx), \min (d^*(\bfx), \piu \od(\bfx)) 
 \end{equation} 
Since the mininum of Lipschitz functions (with the same constant) is 
also Lipschitz (and similarly for maximum), the field $d^{**}$ belongs  also to $\Dn \cap \Lip$. 
The field $d^{**}$ is a better solution than $d^*$ 
because it lowers the objective function:
\begin{equation}\label{key}
	F(d^{**}) \leq F(d^*)
\end{equation}
Indeed,  the objective function $F$ is an integral over $\Ome$
of a convex local function with respect to $d$ denoted $f$ and 
we show that the inequality also holds at the local level 
\begin{equation}\label{key}
	f(d^{**}) \leq f(d^*)
\end{equation}
We have two possible orderings 
\begin{equation}
	d^* < d^{**} = \pil \od \leq \od  \quad 
	\textor \quad \od \leq \piu \od = d^{**} < d^* 
\end{equation}
So, there exist $\lam \in ]0,1[$ such that 
\begin{equation}
	d^{**}  = \lam \od + (1- \lam) d^*
\end{equation}
leading to, by strict convexity of $f$
\begin{equation}
	f(d^{**})  < \lam f(\od) + (1- \lam) f(d^*) = f(d^*) + 
	\lam (f(\od) - f(d^*)) < f(d^*)
\end{equation}
where the last inequality is obtained from the fact that $\od$ at $\bfx$ is the minimum of $f$ at $\bfx$.
%By requiring further strict convexity of $f$ with respect to $d$,
%the last inequality becomes strict.
Thus,  to any field satisfying the constraints but lying outside the bounds, 
we can associate a better one inside the bounds, it proves that the optimal  solution is inside the bounds.

\section{Nonlinear solver}
The nonlinear solver used for \eqref{eq:first_min} 
alternates between a local and global step. The $m$ exponent
is used to denote these iterations.
The local step computes at each integration point 
the stress $\sig^m$ and internal variables 
$\epsp^m, p^m$,   as well as  the tangent operator $T^m$
knowing $u^m, \epsp^n, p^n$ and $d^k$. 
 The $n$ exponent indicates the previous  known time-step 
values whereas the $k$ exponent 
 indicates the latest  computed value of damage  from \eqref{eq:second_min} 
 in the alternate minimization.
 
The global step involves a linear solve which 
finds $u^{m+1} \in U_n$ such that
\begin{equation*}
	\int_0^L T^m \epsilon(u^{m+1}-u^m) \epsilon(u^*) \dint  x =
	- \int_0^L \sig^m \epsilon(u^*) \dint  x, \; \forall 
	u^* \in U_0
\end{equation*}
When the correction $u^{m+1}-u^m$ is below a threshold, we
set  $u^{k+1}, \epsp^{k+1}, p^{k+1}$ as the current $m$ values.
The expressions of $\sig^m$ and $T^m$ are given below for each model.

\subsection{Softening elasticity}
We have
\begin{equation}
	\sig^m = (1- d^k)^2 E \eps^m, \quad \eps^m = \epsilon(u^m), \quad 
	T^m = (1- d^k)^2 E 
\end{equation}
Since the model is linear, a single "$m$" iteration is needed to converge. If dissymmetric tension-compression behavior was considered for damage, more iterations would be needed to converge. 

\subsection{Softening elasticity with hardening plasticity}
The trial stress is computed as 
\begin{equation*}
	\sig^t = E (1-d^k)^2 (\epsilon^m - \epsp^n)
\end{equation*}
followed by the plasticity criterion 
\begin{equation*}
	f^t  = \abs{\sig^t}  - \sigy (1-d^k)^2 (1 + k p^n)
\end{equation*}
giving two cases 
\begin{align*}
	\textif & f^t <= 0  \quad \epsp^{m} = \epsp^{n}, \; p^{m} =  p^{n}, \; \sig^{m} = \sig^t, T^m = E (1-d^k)^2  \\
	\textif & f^t > 0  \quad p^{m} =   \frac{ \abs{\sig^t} -  \sigy + E p^n }{E + \sigy k}, 
	\quad T^m = \frac{E \sigy k (1-d^k)^2}{E + \sigy k} \\
	& \quad \epsp^{m} = \epsp^{n} + (p^{m}-p^n) \frac{\sig^t}{\abs{\sig^t}}, \quad \sig^{m} = E (1-d^k)^2  (\epsilon^m - \epsp^{m})
\end{align*}

\subsection{Softening plasticity}
The trial stress is computed as
\begin{equation*}
	\sig^t = E (\epsilon^m - \epsp^n).
\end{equation*}
followed by the  plasticity criterion  
\begin{equation*}
	f^t  = \abs{\sig^t}  - \sigy (1-d^k)^2 (1 + k p^n)
\end{equation*}
and two cases
\begin{align*}
	\textif & f^t <= 0:  \quad \epsp^{m} = \epsp^{n}, \; p^{m} =  p^{n}, \; \sig^{m} = \sig^t, \; T^m = E \\
	 \textif & f^t > 0: \quad p^{m} =   \frac{ \abs{\sig^t}   -  \sigy (1-d^k)^2 + E p^n}{E + \sigy k (1-d^k)^2}, 
	 \quad T^m = \frac{E \sigy k (1-d^k)^2}{E + \sigy k (1-d^k)^2} \\
	 & \quad \quad \quad \quad \epsp^{m} = \epsp^{n} + (p^{m}-p^n) \frac{\sig^t}{\abs{\sig^t}}, \quad \sig^{m} = E (\epsilon^m - \epsp^{m})
\end{align*}

\bibliographystyle{crunsrt}
\bibliography{refs}

\end{document}